\documentclass[preprint,aps]{revtex4}

\usepackage{graphicx} 
\usepackage{dcolumn}
\usepackage{bm}
\usepackage{color}
\usepackage{tabularx}
\usepackage{doi}

\def\iotabar{\lower3pt\hbox{$\mathchar'26$}\mkern-8mu\iota}

\begin{document}
\bibliographystyle{unsrturl}

\title{The causal impact of magnetic fluctuations in slow and fast L--H transitions at TJ-II}

\author{B.Ph.~van Milligen$^1$, T. Estrada$^1$, B.A. Carreras$^2$, E. Ascas\'ibar$^1$, C. Hidalgo$^1$, I. Pastor$^1$, J.M. Fontdecaba$^1$, R. Balb\'in$^3$ and the TJ-II Team}
 \address{$^1$ Laboratorio Nacional de Fusion, CIEMAT, Avda.~Complutense 40, 28040 Madrid, Spain}
 \address{$^2$ BACV Solutions, 110 Mohawk Road, Oak Ridge, Tennessee 37830, USA}
 \address{$^3$ Instituto Espa{\~n}ol de Oceanograf{\'i}a, Centro Oceanogr{\'a}fico de Baleares, Muelle de Poniente s/n, 07015 Palma de Mallorca, Spain}
\date{\today}

\begin{abstract}
This work focuses on the relationship between L--H (or L--I) transitions and MHD activity in the low {magnetic} shear TJ-II stellarator.
It is shown that the presence of a low order rational {surface} in the plasma edge (gradient) region lowers the threshold density for H--mode access.
MHD activity is systematically suppressed {near} the confinement transition.

We apply a causality detection technique ({based on} the Transfer Entropy) to study the relation between magnetic oscillations and locally measured plasma rotation velocity (related to Zonal Flows). For this purpose, we study a large number of discharges in two magnetic configurations, corresponding to `fast' and `slow' transitions. 
With the `slow' transitions, the developing Zonal Flow prior to the transition {is associated with} the gradual reduction of magnetic oscillations.
The transition itself is marked by a strong spike of `information transfer' from magnetic to velocity oscillations, suggesting that the magnetic drive {may play a role in} setting up the final sheared flow responsible for the H--mode transport barrier.
Similar observations were made for the `fast' transitions.
Thus, it is shown that magnetic oscillations associated with rational surfaces play an important and active role in confinement transitions, 
so that electromagnetic effects should be included in any complete transition model. 
\end{abstract}

\maketitle

\section{Introduction}\label{introduction}

The physics of the Low to High (L--H) confinement transition is a topic of fundamental importance for the operation of fusion experiments and a future fusion reactor.
Since the discovery of the H mode in 1982, the fusion community has devoted an immense effort to improve the understanding of this remarkable phenomenon~\cite{Wagner:2007}, and much progress has been made, although many mysteries remain.
For example, it is known that the L--H transition is affected by magnetic perturbations, although the detailed physics of this process is still unclear.
The external application of 3D magnetic field perturbations (above a certain minimum value) is generally found to raise the L--H power threshold (e.g., at NSTX~\cite{Maingi:2010}, DIII-D~\cite{Gohil:2011}, ASDEX~\cite{Ryter:2013}, MAST~\cite{Scannell:2015}) and to selectively increase particle transport, while hardly affecting energy transport (e.g., at NSTX~\cite{Canik:2010}, DIII-D~\cite{Mordijck:2011}).

Like externally applied magnetic perturbations, internal MHD instabilities are usually considered unfavorable for improved confinement~\cite{Nave:1997}, although simultaneously, it is widely recognized that they may trigger the formation of transport barriers near rational surfaces~\cite{Gruber:2001,Spong:2007}.
Theoretically, the development of Zonal Flows may be facilitated in zones of low Neoclassical shear viscosity near low-order rational surfaces~\cite{Wobig:2000}.
The interaction between MHD modes associated with rational surfaces and Zonal Flows has been addressed in a number of theoretical works~\cite{Scott:2005,Diamond:2005,Grasso:2006,Ishizawa:2007,Ishizawa:2008,Nishimura:2008,Muraglia:2009,Uzawa:2010}.
In summary, under specific circumstances, MHD activity associated with rational surfaces might in fact stimulate the development of Zonal Flows via the Magnetic Reynolds Stress, leading to the formation of transport barriers~\cite{Fujisawa:2009, Estrada:2015}.

MHD activity is often modified sharply at the L--H transition, and usually low frequency MHD modes are suppressed at the transition (e.g., ASDEX~\cite{Toi:1989}, W7-AS~\cite{Wagner:2005}), although they are also sometimes triggered by it (LHD~\cite{Toi:2005,Toi:2009,Solano:2013}). 
Interestingly, MHD fishbone activity has recently  been identified as a trigger of the L--H transition at HL-2A~\cite{Liu:2015}. 
Similarly, at the Globus-M Spherical tokamak, MHD events were sometimes seen to trigger an L--H transition~\cite{Gusev:2008}.
At LHD, an L--H transition was sometimes seen to stabilize MHD modes located well inside the plasma~\cite{Watanabe:2006}.

Experimentally, this issue is most easily addressed at stellarators, as these devices have a better control over the magnetic configuration than, e.g., tokamaks.
Indeed, the dependence of confinement on the magnetic configuration is a well-known feature of stellarators~\cite{Wagner:2005}.
{Significantly}, at W7-AS a dependence of the access to the H-mode on edge iota values was observed~\cite{Wagner:2005}.

This work will specifically address the issue of the interaction of MHD activity and the L--H transition at the TJ-II stellarator.
In order to {study} whether the modification of the MHD activity at the L--H transition is mediated by the modification of profiles or the consequence of a direct interaction between magnetic fluctuations and Zonal Flows, we will make use of a causality detection technique that was recently introduced in the field of plasma physics, {based on} the Transfer Entropy~\cite{Milligen:2015}. 

The Low to High confinement (L--H) transition at TJ-II is a `soft' transition in the sense that the confinement improvement factor is small, and yet it possesses the typical features of any H-mode (rapid drop of $H_\alpha$ emission, reduction of fluctuation amplitudes, formation of a sheared flow layer, ELM-like bursts)~\cite{Tabares:2008,Estrada:2009,Sanchez:2009,Estrada:2010}.
Since the first observation of the H-mode at TJ-II, many experiments have been performed to explore its features.
The present paper will summarize some of this work with special focus on the interaction between the magnetic configuration (rotational transform and Magneto-HydroDynamic or MHD mode activity), L--H transitions and the associated Zonal Flow (ZF).
The currentless, flexible, low {magnetic} shear stellarator TJ-II is ideally suited to study this issue due to complete external control of the magnetic configuration and, in particular, the capacity to modify the rotational transform profile significantly.

The structure of this paper is as follows. 
In Section \ref{statistics}, we will analyze the evolution of some global plasma parameters for a large database of L--H transitions.
In Section \ref{causality}, we analyze the {interactions} between magnetic fluctuations and Zonal Flows during L--H transitions.
Finally, in Section \ref{discussion}, we discuss the results, while in Section \ref{conclusions} we draw some conclusions.

\clearpage
\section{Statistics of L--H transitions at TJ-II}\label{statistics}

The present work summarizes experiments performed between April 2008 and May 2012.
In this period, the H-mode was observed in several hundred discharges, in a range of magnetic configurations.

The TJ-II vacuum magnetic geometry is completely determined by the currents flowing in four external coil sets, meaning that each configuration is fully specified by four numbers. However, the magnetic field is normalized to 0.95 T on the magnetic axis at the ECRH injection point in order to guarantee central absorption of the ECR heating power, so only three independent numbers are left, and these are compounded into a label to identify each configuration~\cite{Milligen:2011c}.
At TJ-II, the normalized pressure $\left < \beta \right >$ is generally low, even in discharges with Neutral Beam Injection (NBI), and currents flowing inside the plasma are generally quite small (unless explicitly driven), so that the actual magnetic configuration is typically rather close to the vacuum magnetic configuration~\cite{Milligen:2011b}.

The experiments have been carried out in pure NBI heated plasmas (line averaged plasma density $\langle n_e \rangle = 2-4 \times 10^{19}$ m$^{-3}$, central electron temperature $T_e = 300-400$ eV). The NBI input heating power is kept constant at about $500$ kW during the discharge but the fraction of NBI absorbed power -- taking into account shine through, CX and ion losses, as estimated using the FAFNER2 code~\cite{Rodriguez:2010} -- increases from 55 to 70\% as the plasma density rises. 

{As noted in the introduction, the L--H transition has a general experimental signature, resulting in a characteristic response of some experimental time traces~\cite{Tabares:2008,Estrada:2009,Sanchez:2009,Estrada:2010}. 
In this work, we specifically define this transition time point as the time at which the amplitude of the density fluctuations measured by the Doppler reflectometer drops sharply. As shown in Ref.~\cite{Estrada:2009}, this time is more precise than, e.g., the time needed for the formation of the shear layer, which may take several ms. 
This procedure typically allows identifying the L--H transition time with a precision of about 1 ms. 
}

\begin{figure}\centering
  \includegraphics[trim=100 50 100 0,clip=,width=12cm]{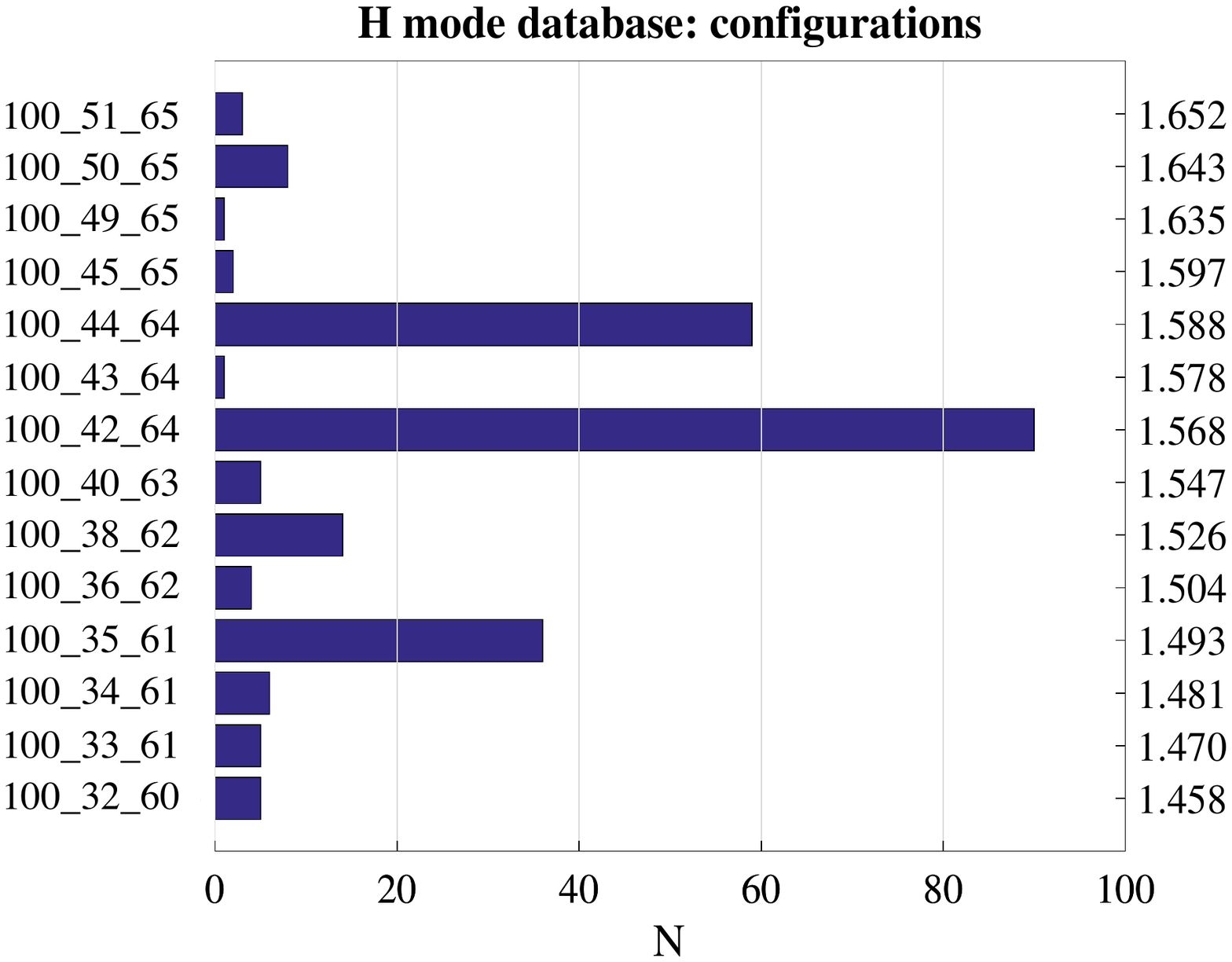}
\caption{\label{Config} Configuration breakdown of L--H transition discharges at TJ-II ($N$ is the number of discharges).
The right axis specifies the value of $\iotabar$ at $\rho=2/3$.}
\end{figure}

\begin{figure}\centering
  \includegraphics[trim=0 0 0 0,clip=,width=16cm]{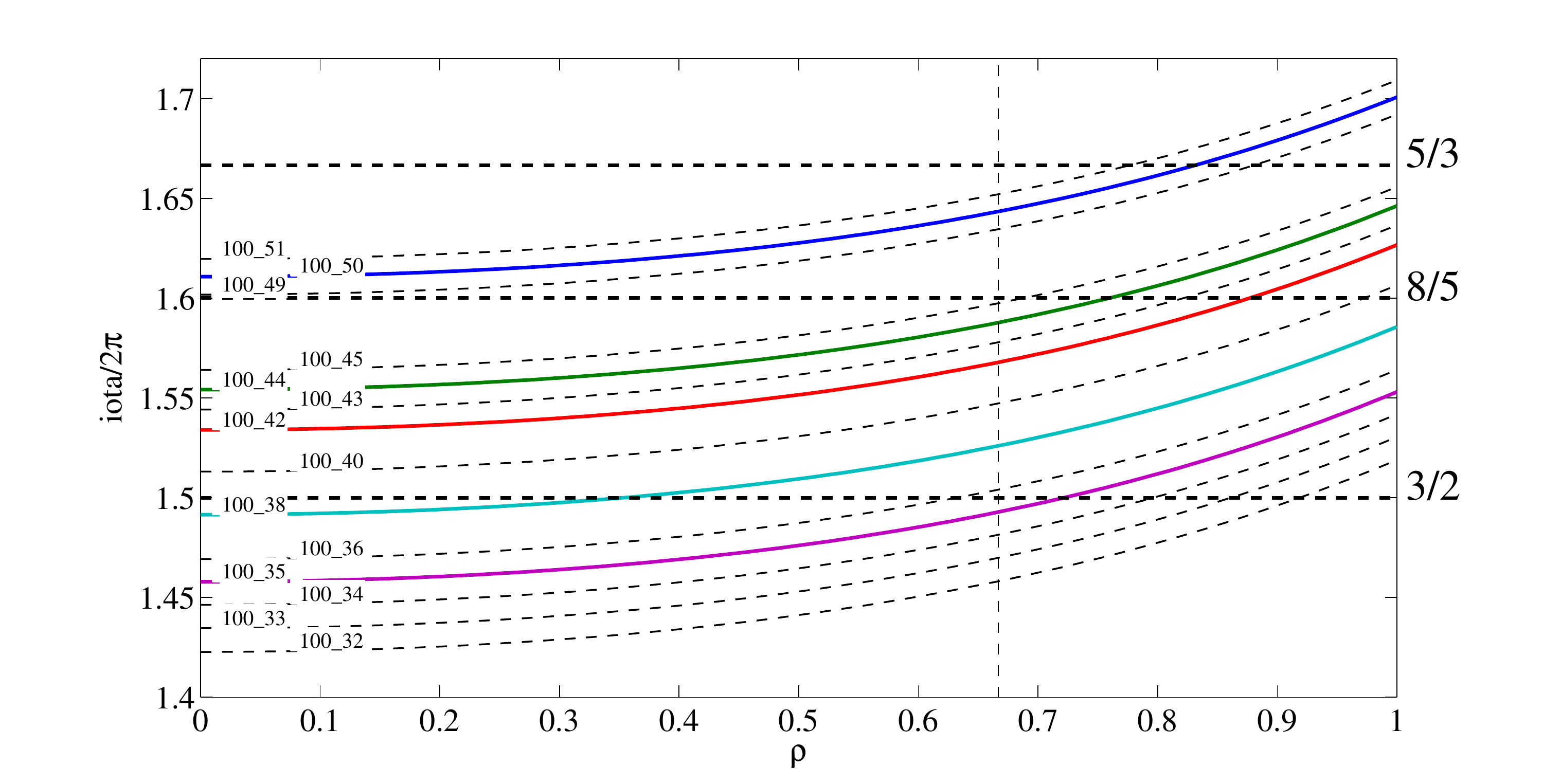}
\caption{\label{iota} Vacuum rotational transform of the magnetic configurations considered (the line labels specify the first two numbers of the configuration label). The rotational transforms of the most common configurations are drawn as continuous curves, those of the remaining configurations are dashed. Horizontal dashed lines indicate the main low-order rational values. The vertical dashed line indicates the position $\rho = 2/3$.}
\end{figure}

Fig.~\ref{Config} shows the breakdown of observed L--H transitions according to magnetic configuration in a database of 218 discharges.
Fig.~\ref{iota} shows the vacuum rotational transform of the magnetic configurations considered.
For simplicity, the magnetic configurations considered here are also identified by the value of the vacuum rotational transform ($\iotabar = \iota/2\pi = 1/q$) at $\rho = r/a = 2/3$, which is immediately inward from the gradient region for most discharges~\cite{Milligen:2011b}.

\begin{figure}\centering
  \includegraphics[trim=0 0 0 0,clip=,width=12cm]{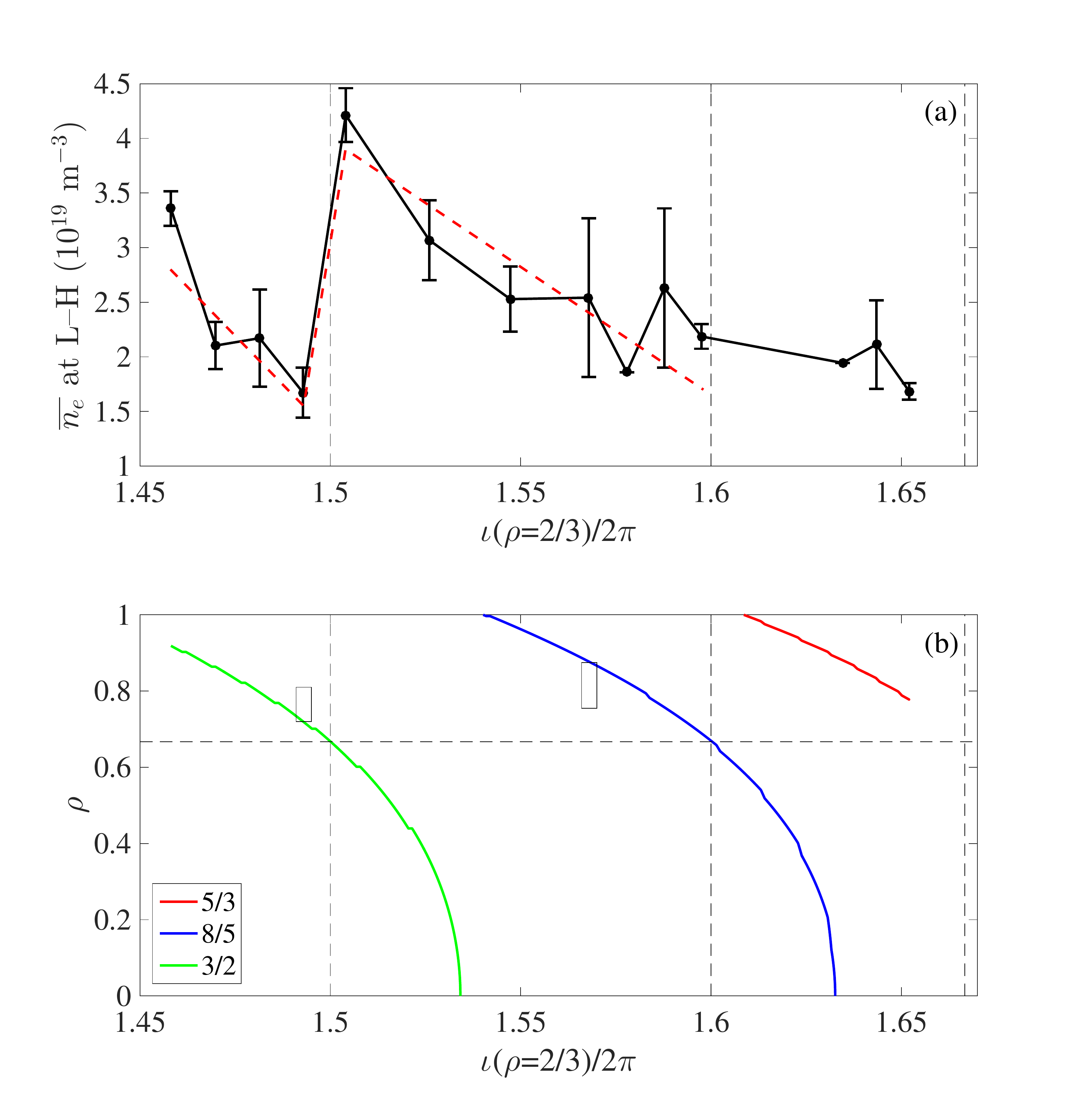}
\caption{\label{Config_all} Magnetic configuration scan (configurations identified via $\iotabar(\rho=2/3)$). 
(a) Line average density at L--H transition. The red dashed line is provided to guide the eye. The bars indicate the shot to shot variation (not the measurement error). 
(b) Position of the main low order rational surfaces in vacuum. The horizontal dashed line indicates $\rho=2/3$.
The vertical dashed lines correspond to $\iotabar(\rho=2/3) = 3/2, 8/5,$ and 5/3. 
The two small rectangles indicate the approximate measurement locations of Doppler Reflectometry, as discussed in the text.
}
\end{figure}

Fig.~\ref{Config_all}a shows the mean line average density at the L--H transition.
Fig.~\ref{Config_all}b shows the position of the main low order rational surfaces.
The figure shows that the line average density at which the L--H transition occurs is lower when a major low-order rational exists in the outer regions of the plasma, suggesting that a low-order rational in the edge region facilitates the L--H transition.

\clearpage
\subsection{Slow and fast L--H transitions}

In this work, we will mainly focus on two magnetic configurations { with good statistics and Doppler Reflectometry data (cf.~Fig~\ref{Config}, noting that configuration 100\_44\_64 behaves very similar to 100\_42\_64)}:
\begin{itemize}
\item In configuration 100\_35\_61, with $\iotabar(\rho=2/3) = 1.493$, the radial position of the $\iotabar = 3/2$ rational is located at $\rho \simeq 0.73$ in vacuum, cf.~Fig.~\ref{iota}. This configuration is characterized by a `slow' transition. The transition is not straight into the H phase, but rather into an I phase, characterized by Limit Cycle Oscillations (LCOs), as reported elsewhere~\cite{Estrada:2011,Estrada:2015} and similar to LCOs reported at other devices~\cite{Colchin:2002}.
\item In configuration 100\_42\_64, with $\iotabar(\rho=2/3) = 1.568$, the radial position of the $\iotabar = 8/5$ rational is located at $\rho \simeq 0.86$ in vacuum, cf.~Fig.~\ref{iota}. In this case, the transition is `fast' and enters directly into the H phase.
\end{itemize}

\begin{figure}\centering
  \includegraphics[trim=0 0 200 0,clip=,width=14cm]{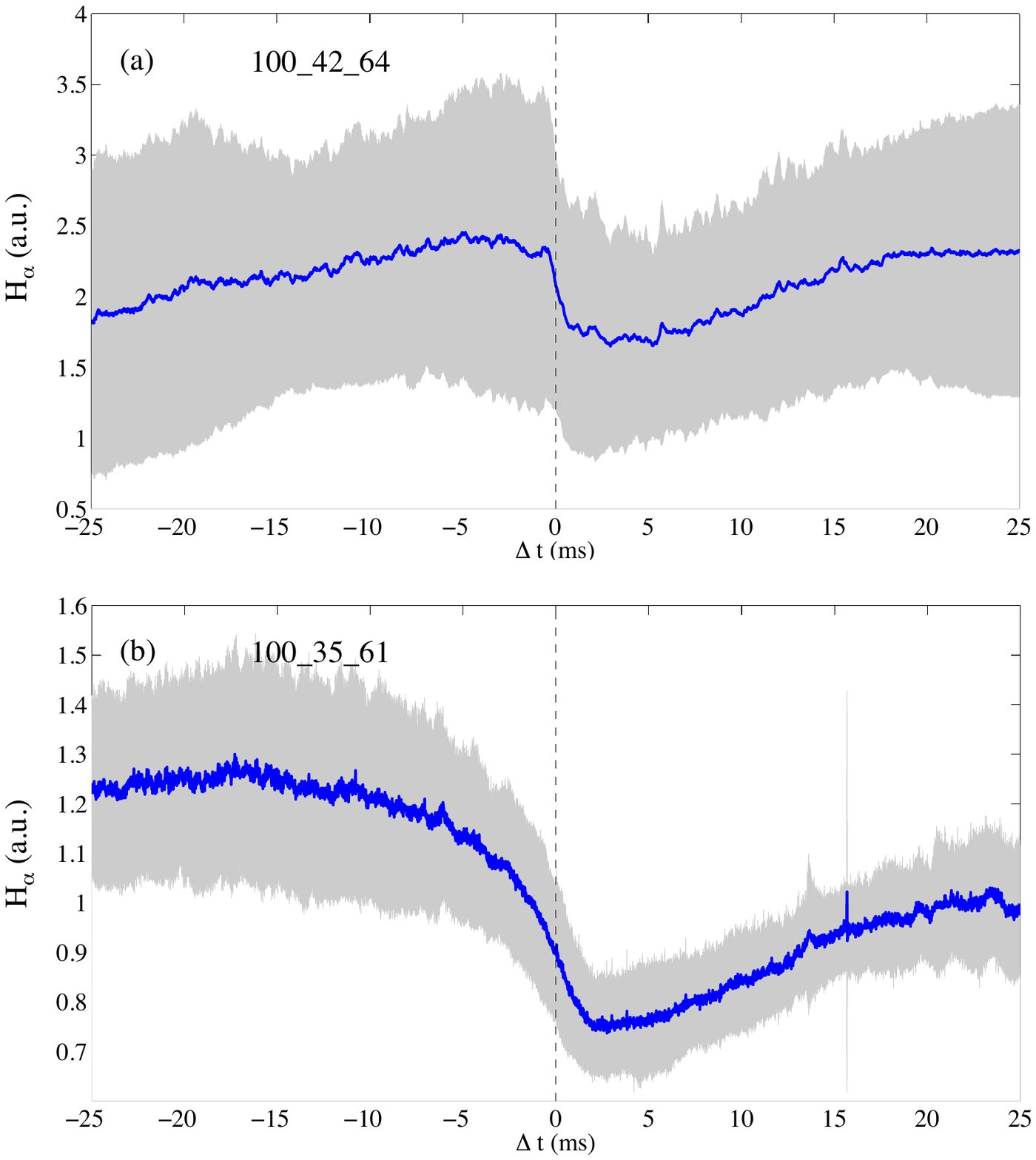}
\caption{\label{halpha}Mean evolution of $H_\alpha$ emission across confinement transitions. 
(a): configuration 100\_42\_64 (`fast' L--H transitions). 
(b): configuration 100\_35\_61 (`slow' L--I transitions).}
\end{figure}

Fig.~\ref{halpha} shows the evolution of a specific $H_\alpha$ monitor (noting that other $H_\alpha$ signals at other toroidal locations behave similarly), averaged over the discharges belonging to each configuration, versus $\Delta t$, where $\Delta t$ is the time minus the L--H transition time in ms.
In this and the succeeding graphics, the grey area indicates the shot to shot variation (not the error).
The drop in $H_\alpha$ is one of the markers used to identify such transitions.

\begin{figure}\centering
  \includegraphics[trim=0 0 0 0,clip=,width=14cm]{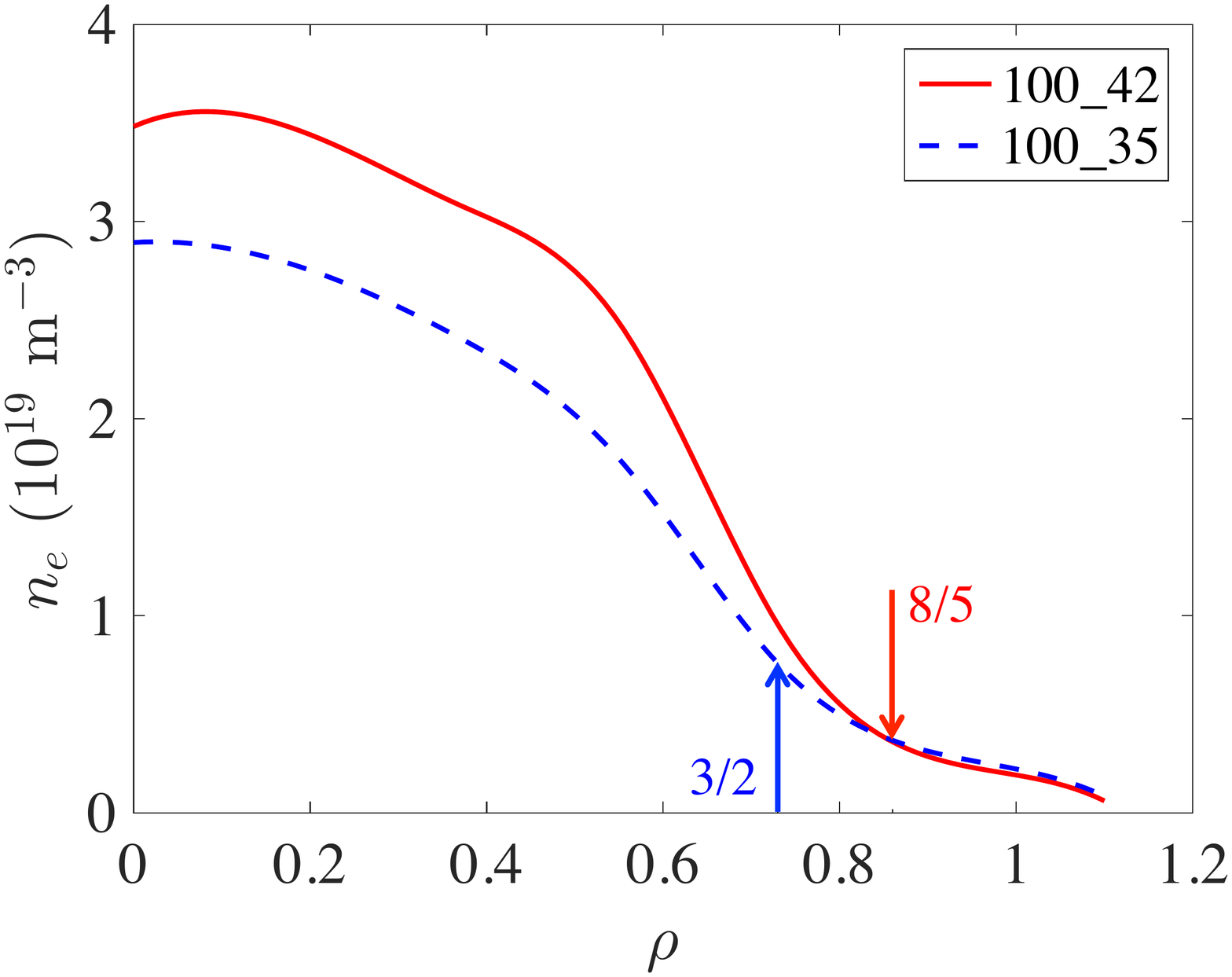}
\caption{\label{ne_profiles}Mean density profiles at the L--H transition time, calculated using a Bayesian technique and averaged over 32 and 26 discharges (for configuration 100\_42\_64, `fast' L-H transitions, and 
configuration 100\_35\_61, `slow' L-I transitions, respectively).
Arrows indicate the (vacuum) location of the main low order rational {surface} in the edge for each configuration.
}
\end{figure}

Fig.~\ref{ne_profiles} shows the mean density profiles at the time of the L--H transition, calculated using a Bayesian profile reconstruction technique described elsewhere~\cite{Milligen:2011b}, and averaged over a number of discharges.
The location of the main low-order rational { surface} in the edge region is indicated, showing that the rational { surface} is located near the `foot' of the gradient region in both cases.

\clearpage
\subsection{Evolution of magnetic activity during L--H transitions}

\begin{figure}\centering
  \includegraphics[trim=0 0 0 0,clip=,width=16cm]{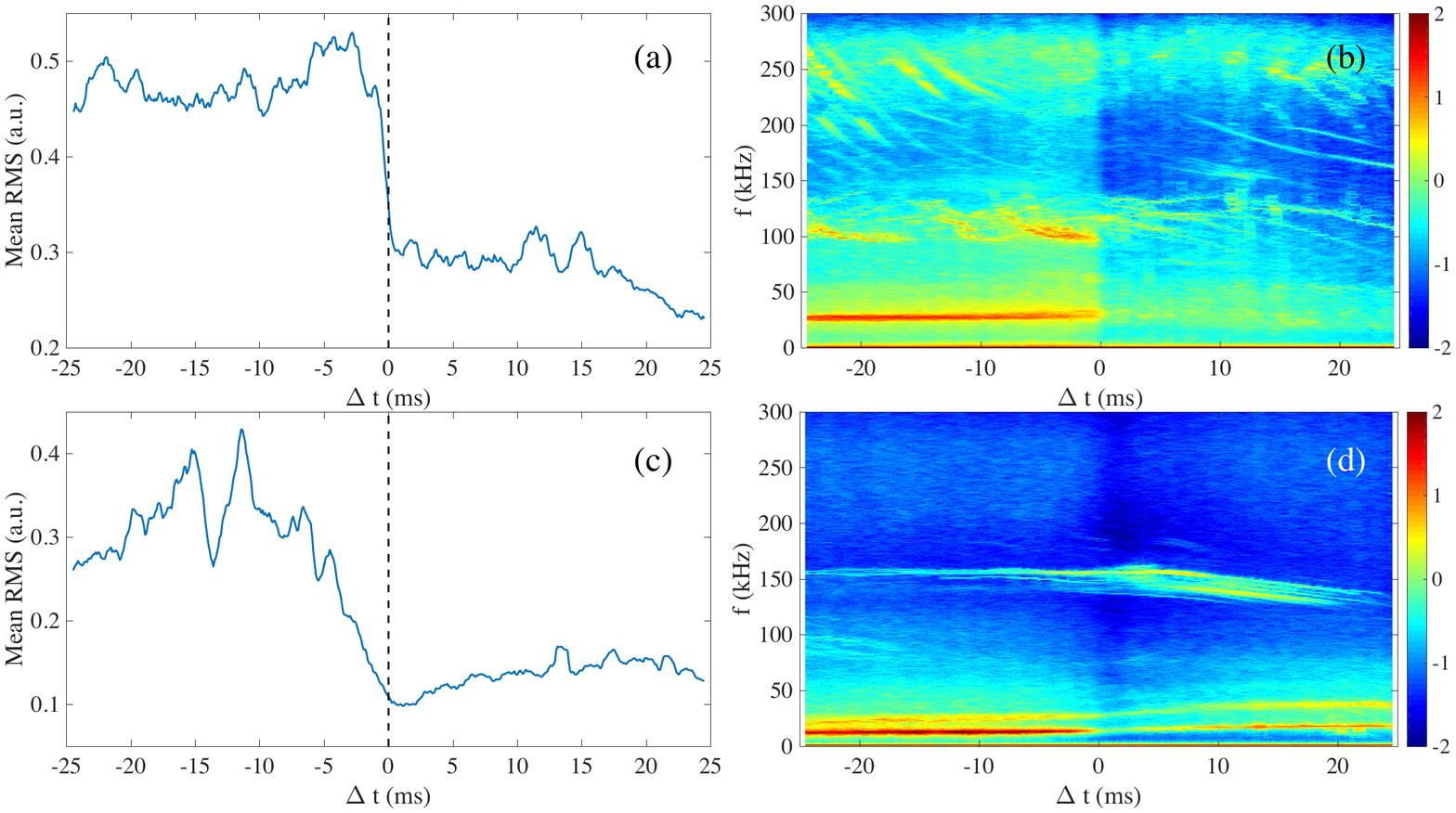}
\caption{\label{mirnov}Mean evolution of magnetic activity across confinement transitions. 
(a,b): configuration 100\_42\_64 (`fast' L--H transitions, 90 discharges). 
(c,d): configuration 100\_35\_61 (`slow' L--I transitions, 36 discharges).
(a,c): mean RMS of a Mirnov coil. 
(b,d): mean spectrogram of a Mirnov coil.}
\end{figure}

We take the Root Mean Square (RMS) amplitude of a specific Mirnov coil signal as a global measure for MHD activity.
We calculate the RMS of the coil signal using 0.2 ms overlapping time bins.
Fig.~\ref{mirnov} shows both the RMS evolution of the coil signal and the mean spectrum versus time for this coil.
In the case of the `slow' transitions, one observes a gradual decrease of MHD activity prior to the L--H transition {time (as defined in the manner described above)}, the decay starting some ten ms beforehand.
{This early decay with respect to the reference time is also visible, to some degree, in the $H_\alpha$ trace shown in Fig.~\ref{halpha}b.}
In the case of the `fast' transitions, the drop of RMS is rather sharp and lasts only a few ms, although one could still argue that the decay starts before the transition time.
The drop in RMS of the Mirnov coil signal from the pre-transition peak to the transition point is typically by a factor of about 2.
Afterwards, in I or H mode, {the RMS amplitude} remains low.
The spectra show that the drop of RMS is mainly associated with the suppression of low frequency ($f < 80$ kHz) MHD modes, and not so much with that of high frequency ($f > 80$ kHz) Alfv\'en modes.

Using a poloidal set of Mirnov coils~\cite{Jimenez:2011}, it is possible to identify the dominant mode numbers of the low frequency modes before the L--H transition, 
leading to the result shown in Table \ref{table2}.
In each case, the dominant mode number $m$ corresponds to the main low order rational {surface} present inside the plasma according to the
vacuum rotational transform profile, cf.~Fig.~\ref{iota}.
The dominant frequency of the corresponding mode is indicated, and when a given mode (8/5 or 3/2) is further inward, {the} frequency is higher due to a higher rotation velocity.
{ This is in accordance with earlier work, which showed} the existence of a rotation velocity profile that increases from the edge inward~\cite{Milligen:2011c}.
\begin{table}[htp]
\caption{Dominant mode numbers prior to L--H transition}
\begin{center}
\begin{tabular}{cccccc}
\hline
Configuration	& $\iotabar(\rho=2/3)$	& Dominant mode	&Lowest rational 	& $\rho$ (rational) 	& Frequency \\
			& 				& number, $m$ 	&				& 				& (kHz) \\
\hline
100\_50\_65	& 1.643	&3	&5/3	&	0.83	 	&17\\
100\_44\_64	& 1.588	&5	&8/5	&	0.76	 	&60\\
100\_42\_64	& 1.568	&5	&8/5	&	0.86 		&27\\
100\_38\_62	& 1.526	&2 	&3/2	&	0.33 		&$\sim 80$\\
100\_35\_61	& 1.493	&2	&3/2	&	0.73 		&12\\
\hline
\end{tabular}
\end{center}
\label{table2}
\end{table}%

{The decay} of magnetic activity {associated with} confinement transitions (Fig.~\ref{mirnov}) is rather striking.
A priori, the origin of this phenomenon is unclear.
{One might be inclined to think} that the decay is related to a gradual evolution of profiles on the transport time scale (of the order of 5 ms) associated with the rising density (cf.~ Fig.~\ref{density}).
However, in the case of the `fast' L--H transition (configuration 100\_42\_64), the decay is sufficiently fast to make it doubtful that transport effects are playing a significant role.
{Section \ref{causality}} will attempt to {shed some light on} the origin of this phenomenon.

\subsection{Temporal evolution of some global parameters}

\begin{figure}\centering
  \includegraphics[trim=0 0 200 0,clip=,width=14cm]{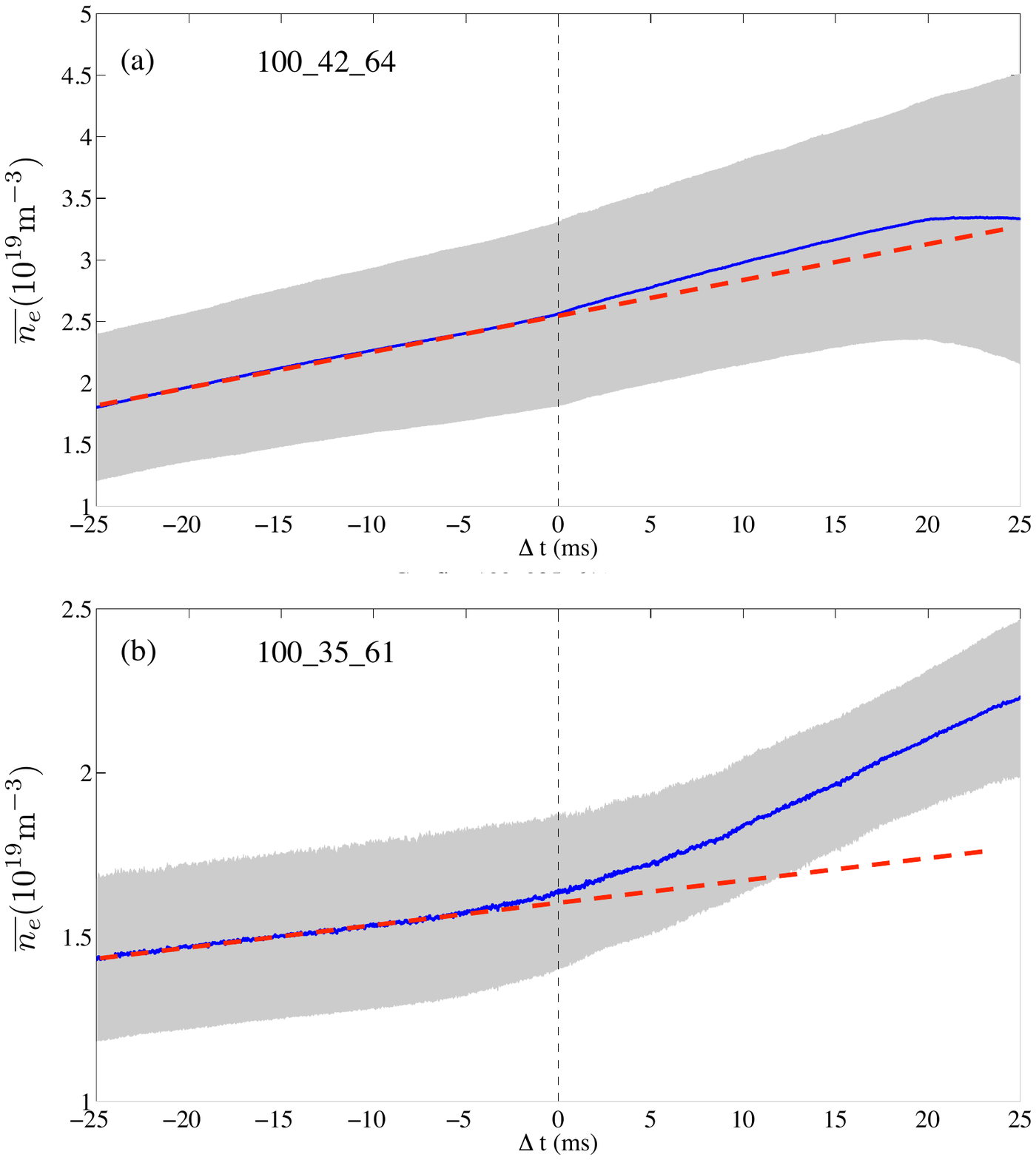}
\caption{\label{density}Mean evolution of the line average density across confinement transitions (units: $10^{19}$ m$^{-3}$).
(a): configuration 100\_42\_64 (`fast' L--H transitions). 
(b): configuration 100\_35\_61 (`slow' L--I transitions).
{Dashed red lines indicate a linear extrapolation of the evolution prior to $\Delta t = 0$.}}
\end{figure}

Fig.~\ref{density} shows the evolution of the mean line average density $\overline{n_e}$ as measured by the interferometer~\cite{Milligen:2011b}.
It should be borne in mind that the mean temporal evolution away from the transition time may be affected by irrelevant events, such as ECRH switch-off or NBI switch-on for large negative values of $\Delta t$, and plasma termination for large positive values of $\Delta t$
{(this argument applies to all graphs of this paper having  $\Delta t$ on the abscissa).}
However, one observes that the time derivative of the density evolution is increased slightly at the transition, associated with an enhancement of particle confinement.

\begin{figure}\centering
  \includegraphics[trim=0 0 200 0,clip=,width=14cm]{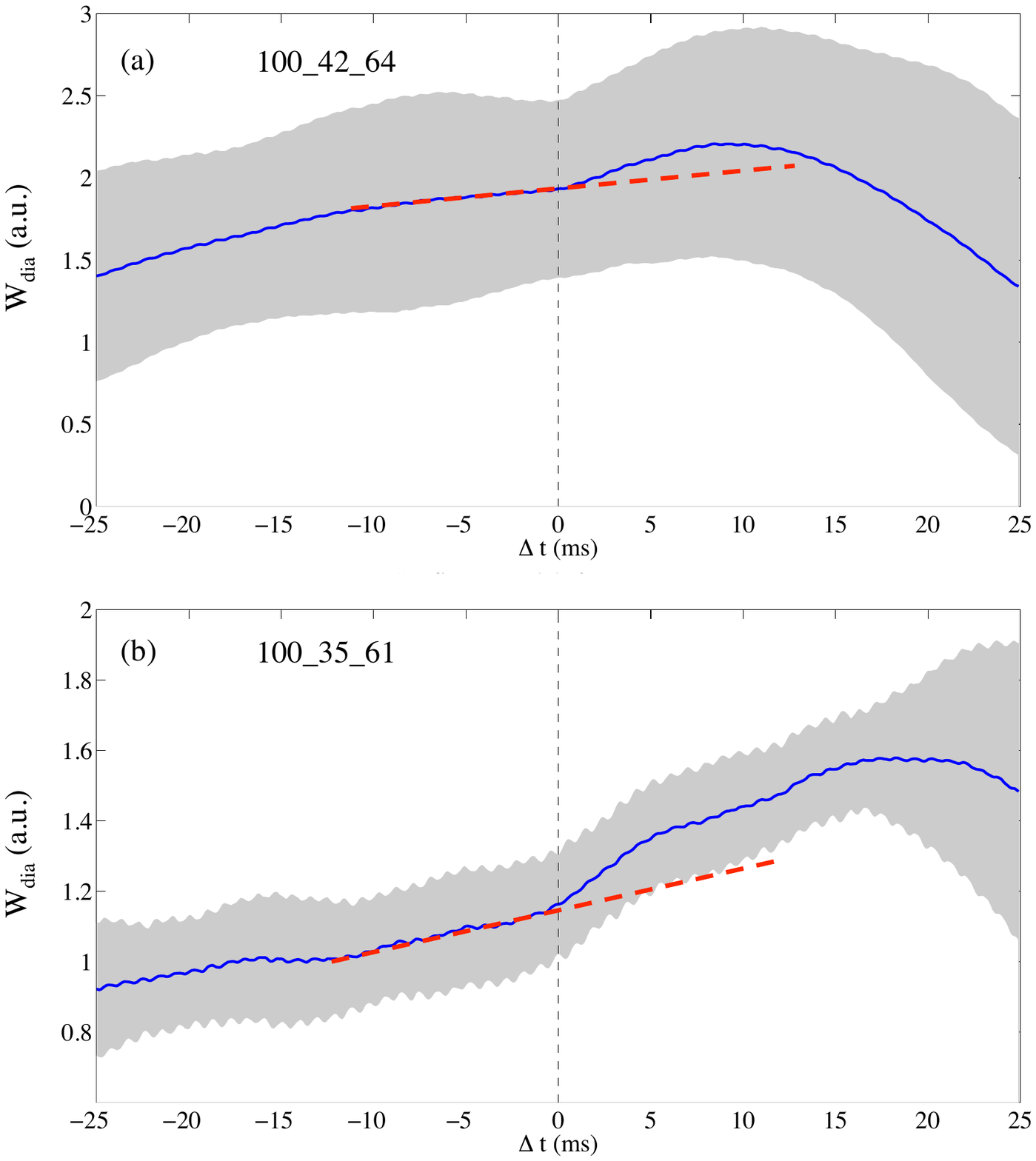}
\caption{\label{wdia}Mean evolution of $W_{\rm dia}$ across confinement transitions.
(a): configuration 100\_42\_64 (`fast' L--H transitions). 
(b): configuration 100\_35\_61 (`slow' L--I transitions).
{Dashed red lines indicate a linear extrapolation of the evolution prior to $\Delta t = 0$.}}
\end{figure}

Fig.~\ref{wdia} shows a summary of the evolution of the corrected diamagnetic energy content~\cite{Ascasibar:2010}.
In {both} cases, a significant change of slope is visible, coincident {(within about a ms)} with the L--H transition time {(as defined above)}.
Since the energy confinement time is defined as $\tau_E = W_{\rm dia}/(P_{\rm abs}-dW_{\rm dia}/dt)$, where $P_{\rm abs}$ is the absorbed power, this change of slope implies a change in energy confinement time.
{The sharpness of the change in slope implies that our definition of the L--H transition time is indeed reliable to about one ms.}

\begin{figure}\centering
  \includegraphics[trim=0 0 0 0,clip=,width=16cm]{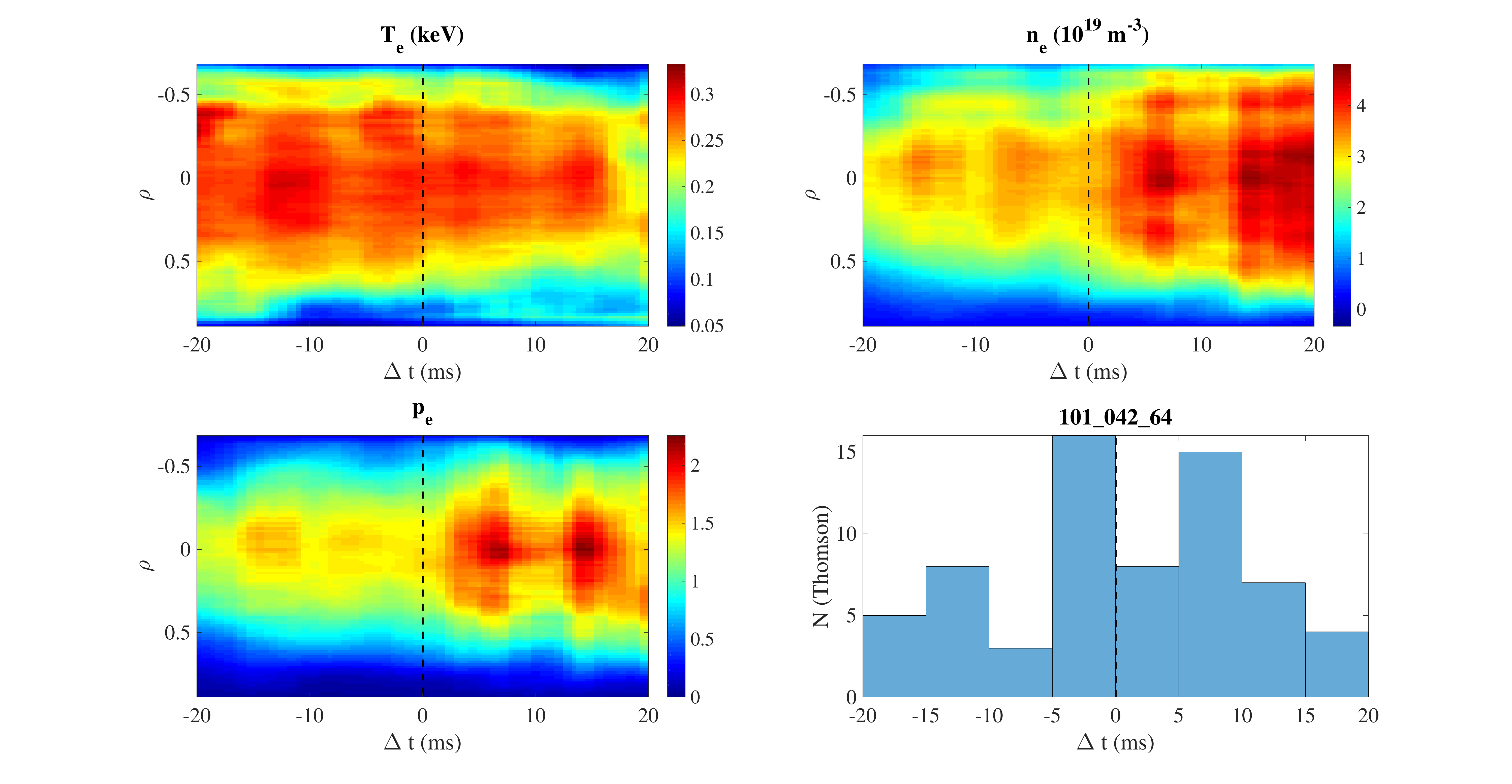}\
\caption{\label{TS}Evolution of mean Thomson Scattering data across the L--H transition in configuration 100\_42\_64 (`fast' transitions).
Top left: $T_e$; top right: $n_e$; bottom left: $p_e$; bottom right: number of Thomson Scattering profiles per 5 ms bin.}
\end{figure}

Fig.~\ref{TS} shows the mean evolution of the Thomson Scattering (TS)~\cite{Barth:1999} profiles across the L-H transition in configuration 100\_42\_64 (the most populated case, corresponding to the `fast' transition; { insufficient data were available for the `slow' transitions to produce a similar graph}). The TJ-II TS diagnostic is a single pulse diagnostic, so the evolution shown is obtained by 
{interpolating between profiles measured in} different discharges {at different times}, each {discharge being characterized by} slightly different densities and temperatures. 
At high densities, the temperature tends to drop due to radiation effects~\cite{Tabares:2010}.
{In spite of the fact that the discharges are similar but not identical}, the mean evolution is rather clear: across the transition, the $T_e$ profile is mostly constant or slightly decreasing in amplitude; on the other hand, the $n_e$ profile increases significantly in amplitude and width, associated with the establishment of the edge transport barrier (in accordance with results reported previously~\cite{Milligen:2011b}).
Thus, the edge transport barrier mainly affects the electron density, and hardly affects the electron temperature.

\begin{figure}\centering
  \includegraphics[trim=0 0 0 0,clip=,width=12cm]{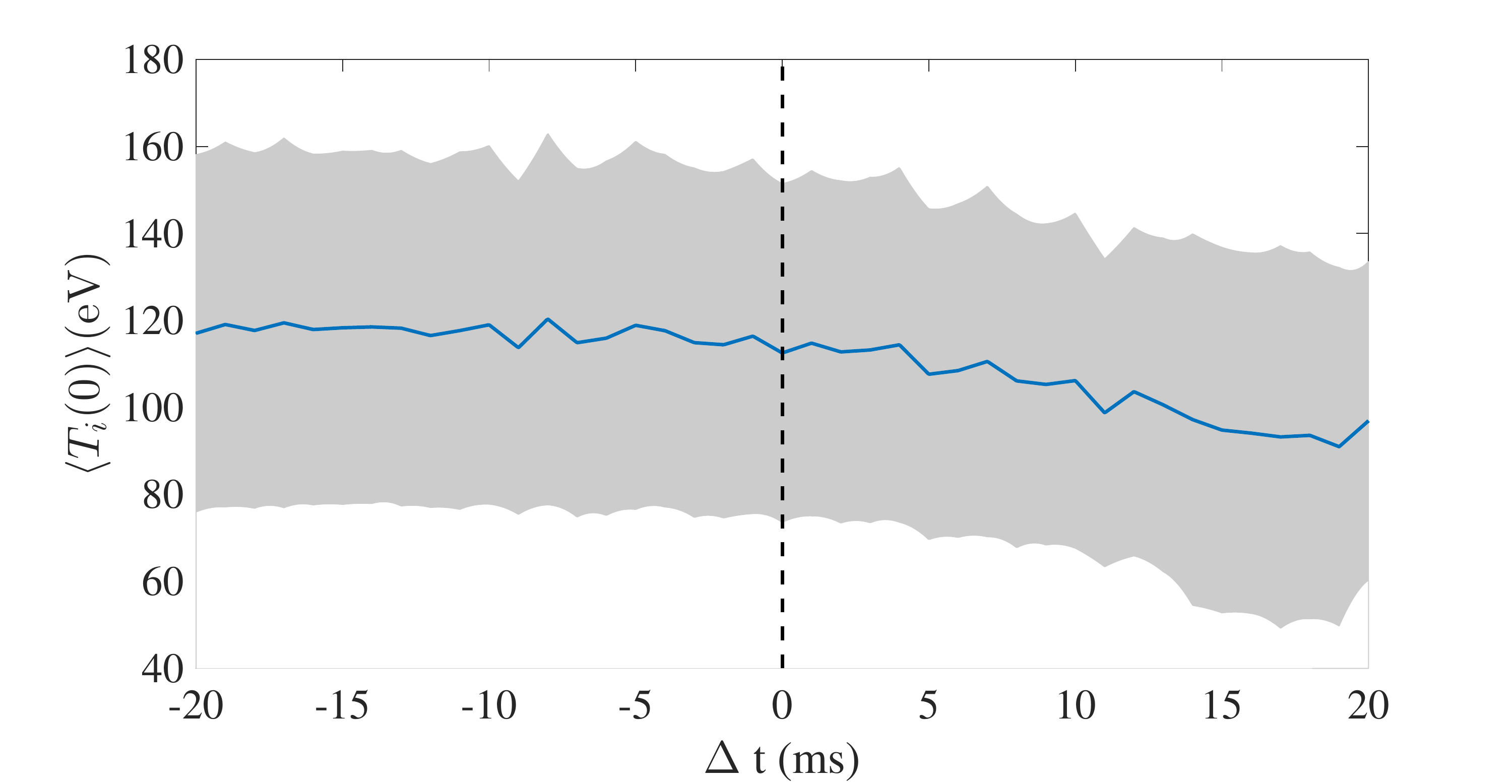}\
\caption{\label{Ti}{The evolution of $\langle T_i(0)\rangle$, the average of the core ion temperature $T_i(0)$ over 20 discharges,} across the L--H transition in configuration 100\_42\_64 (`fast' transitions). 
The grey area {indicates} the standard deviation of the shot to shot variation (not the error).}
\end{figure}

Fig.~\ref{Ti} shows the mean evolution of the core ion temperature, measured by the charge exchange diagnostic~\cite{Fontdecaba:2014} { in configuration 100\_42\_64 (`fast' transitions); insufficient data were available for the `slow' transitions to produce a similar graph}.
The {core} ion temperature is shown to decrease slightly after the L--H transition, similar to $T_e$.
Thus, the observed increase of $\tau_E$ is due to the increase in density. 

\clearpage
\section{Interaction between magnetic and Zonal Flow oscillations}\label{causality}

The previous section has shown that the magnetic configuration plays a role in confinement transitions at TJ-II (cf.~Fig.~\ref{Config_all}),
but it does not clarify the details.
It is known that Zonal Flows play a major role in the confinement transitions in general~\cite{Diamond:2005}, and this is also the case at TJ-II~{\cite{Hidalgo:2009,Pedrosa:2010}}.
To study the interaction between magnetic fluctuations and Zonal Flows, we will use the signal from a magnetic poloidal field pick-up coil ($\dot B$) and signals from the Doppler reflectometry system installed at TJ-II.
Although a Mirnov coil measures global magnetic activity, its signal is often dominated by low-order modes associated with low-order rational surfaces, and we exploit this property here.
The Doppler reflectometer allows measuring the perpendicular rotation $v_\perp$ (within the flux surface) of fluctuations at two specific radial locations~\cite{Happel:2009}.
The perpendicular velocity, $v_\perp(\rho,t)$, is calculated with high temporal resolution~\cite{Estrada:2012}.

To analyze the mentioned interaction, we apply a causality detection technique~\cite{Schreiber:2000,Hlavackova:2007,Milligen:2014}.
The Transfer Entropy between signals $Y$ and $X$ quantifies the number of bits by which the prediction of a signal $X$ can be improved by using the time history of not only the signal $X$ itself, but also that of signal $Y$ (Wiener's `quantifiable causality').

Consider two processes $X$ and $Y$ yielding discretely sampled time series data $x_i$ and $y_j$.
In this work, we use a simplified version of the Transfer Entropy,
\begin{equation}\label{TE}
T_{Y \to X} = \sum{p(x_{n+1},x_{n-k},y_{n-k}) \log_2 \frac{p(x_{n+1}|x_{n-k},y_{n-k})}{p(x_{n+1}|x_{n-k})}}
\end{equation}
{Here, $p(a,b,c)$ is a joint probability distribution over the variables $a,b$ and $c$, while $p(a|b)$ is a conditional probability distribution, $p(a|b)=p(a,b)/p(b)$.}
The sum runs over the arguments of the probability distributions (or the corresponding discrete bins {when the probability distributions are constructed using data binning}).
The number $k$ is the `time lag index' shown in the graphs in the following sections.
The construction of the probability distributions is done using `course graining', i.e., a low number of bins (here, $m=5$), to obtain statistically significant results. 
The value of the Transfer Entropy $T$, expressed in bits, can be compared with the total bit range, $\log_2 m$, equal to the maximum possible value of $T$, to help decide whether the transfer entropy is significant or not.

{A simple way of estimating the statistical significance of the Transfer Entropy is by calculating $T$ for two random (noise) signals.
Fig.~\ref{TE_random} shows the Transfer Entropy calculated for two such random signals, with a Gaussian distribution, each with a number of samples equal to $N$.
It can be seen that the value of the Transfer Entropy (averaged over 100 equivalent realizations) drops proportionally to $1/N$.

\begin{figure}\centering
  \includegraphics[trim=0 0 0 0,clip=,width=12cm]{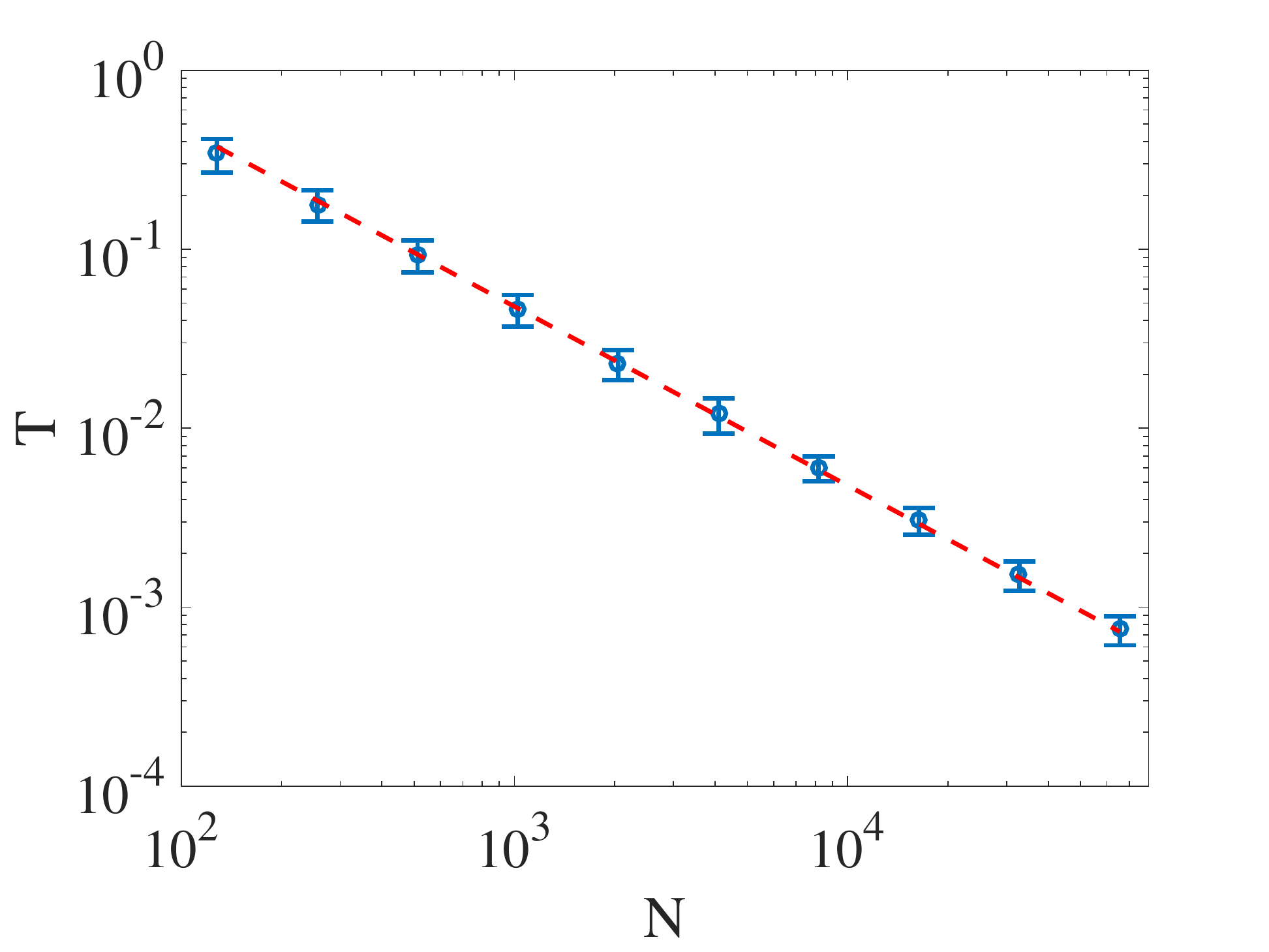}
\caption{\label{TE_random}{Transfer Entropy for two random Gaussian signals, as a function of the number of samples of the signals, $N$ (with $m=5$). 
Each point is calculated as the average over 100 independent realizations, and the error bar indicates the variation of the result.
This average value can be taken as the statistical significance level of the Transfer Entropy. The red dashed line is proportional to $1/N$.}}
\end{figure}

}

In interpreting the Transfer Entropy, it should be noted that it is a non-linear quantifier of information transfer that can help clarifying which fluctuating variables influence which others. In this sense, it is fundamentally different from the (linear) cross correlation, which only measures signal similarity. For example, the cross correlation is maximal for two identical signals ($X=Y$), whereas the Transfer Entropy is exactly zero for two identical signals (as no information is gained by using the second, identical signal to help predicting the behavior of the first).

As with all methods for causality detection, an important caveat is due. The method only detects the information transfer between measured variables. If the net information flow suggests a causal link between two such variables, this may either be due to a direct cause/effect relation, or due to the presence of a third, undetected variable that can mediate or influence the information flow.

\begin{figure}\centering
  \includegraphics[trim=0 100 200 0,clip=,width=16cm]{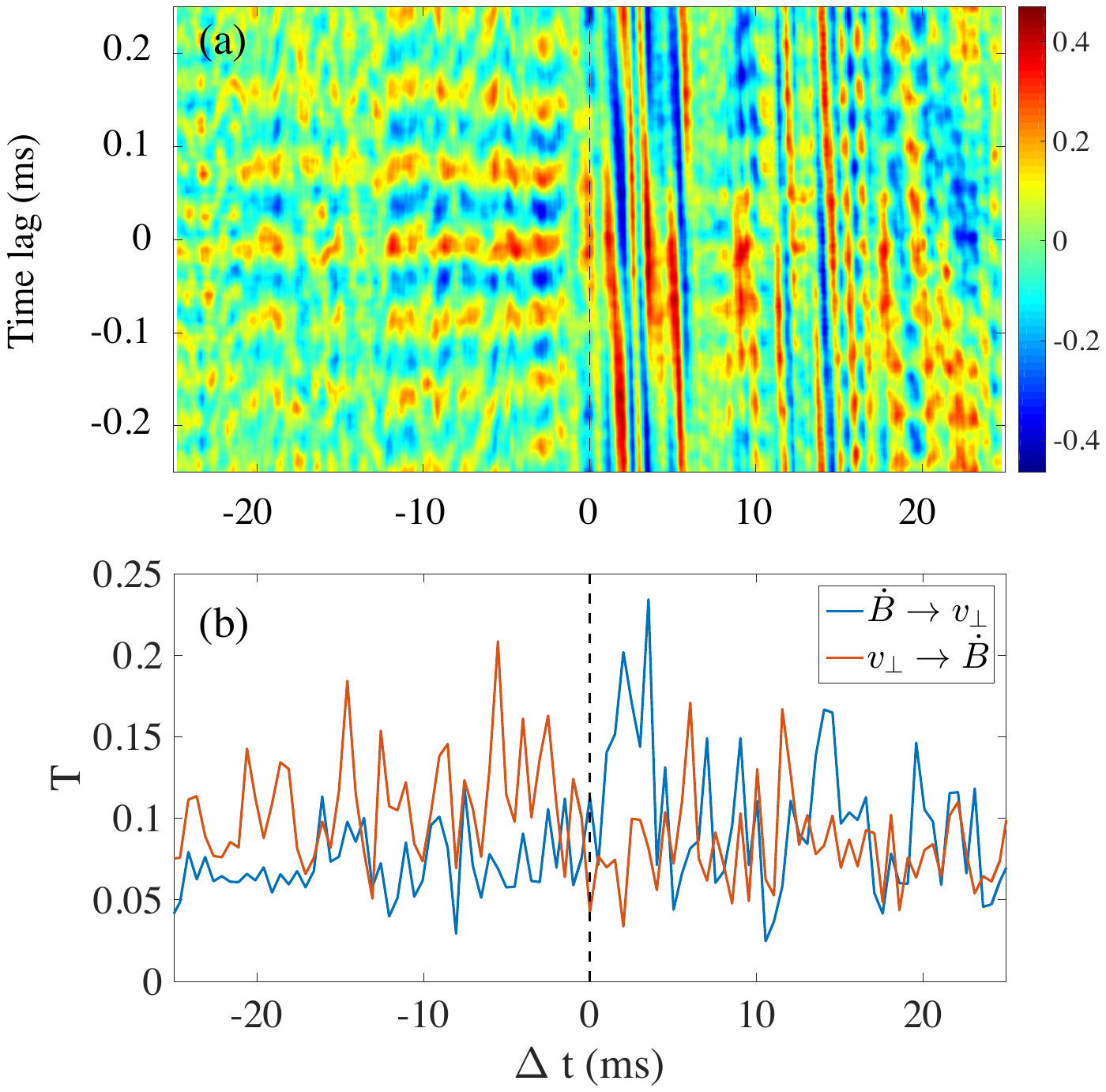}
\caption{\label{27135}Discharge 27135, configuration 100\_35\_61 ($\iotabar(\rho=2/3) = 1.493$, {`slow'}). 
(a): Correlation between $\tilde v_\perp$ of Doppler reflectometry channel 1 ($\rho \simeq 0.77$ in the H phase) and Mirnov coil signal, $\dot B_\theta$ (vertical axis: time delay).
(b): Transfer Entropy between $\tilde v_\perp$ of Doppler reflectometry channel 1 and a Mirnov coil signal, $\dot B_\theta$. Settings: $m=5$, lag $k=500$, or 50 $\mu$s.}
\end{figure}

Fig.~\ref{27135} (a) shows the linear correlation between $v_\perp$ and a Mirnov coil signal, $\dot B_\theta$, in configuration 100\_35\_61 (`slow' transitions).
The correlation has been calculated using overlapping time windows with a length of 0.5 ms each.
Before the transition, a regular structure is visible, mainly reflecting the presence of an MHD mode with a frequency of slightly more than 10 kHz.
Immediately after the transition, lasting slightly more than 5 ms, a relatively strong correlation is visible as a diagonal striped feature.

Fig.~\ref{27135} (b) shows the Transfer Entropy, calculated with $m=5$, using overlapping time windows of 0.5 ms, for a lag of $k=500$, equivalent to 50 $\mu$s.
{The 0.5 $\mu$s time window corresponds to $N=5000$ data points, so the statistical error level of the Transfer Entropy is $\simeq 0.01$ (cf.~Fig.~\ref{TE_random}).}
As noted above, the theoretical maximum of $T$ equals $\log_2 m = 2.3$; reported values can be compared to this maximum.
The graph shows that following the transition time, the magnetic signal contains significant information to predict the evolution of $v_\perp$ on a time scale of 50 $\mu$s.
Note the coincidence of this time window with the correlation signature seen in Fig.~\ref{27135} (top).

\begin{figure}\centering
  \includegraphics[trim=0 100 200 0,clip=,width=16cm]{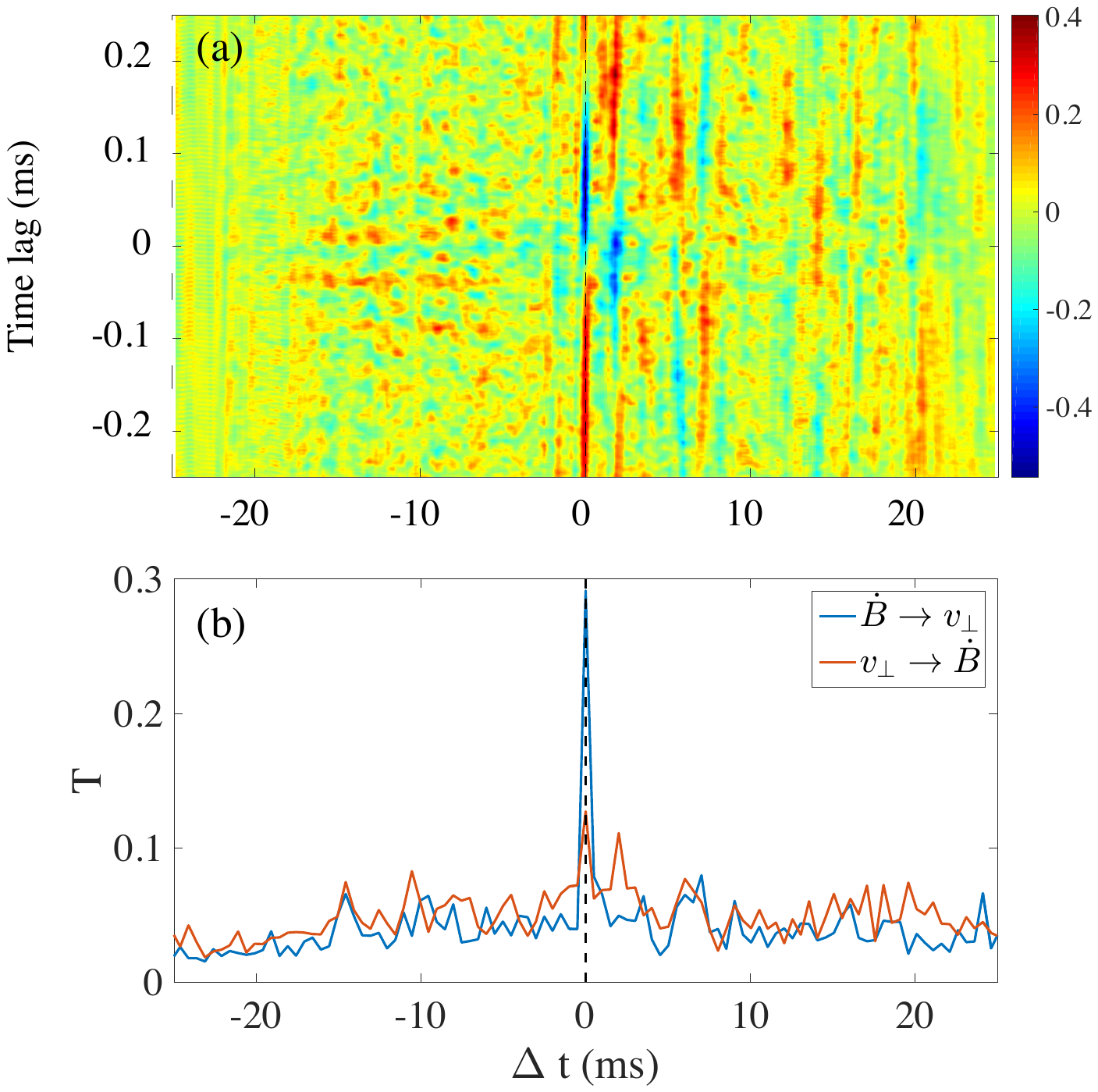}
\caption{\label{23029}Discharge 23029, configuration 100\_42\_64 ($\iotabar(\rho=2/3) = 1.568$, {`fast'}). 
(a): Correlation between $\tilde v_\perp$ of Doppler reflectometry channel 1 ($\rho \simeq 0.86$ in the H phase) and Mirnov coil signal, $\dot B_\theta$ (vertical axis: time delay).
(b): Transfer Entropy between $\tilde v_\perp$ of Doppler reflectometry channel 1 and a Mirnov coil signal, $\dot B_\theta$. Settings: $m=5$, lag $k=500$, or 50 $\mu$s.}
\end{figure}

Fig.~\ref{23029} shows similar results for configuration 100\_42\_64 (`fast' transitions).
Again, a vertically striped structure appears in the correlation at the transition.
The Transfer Entropy peaks sharply at the transition, the dominant {direction of the interaction} being from $\dot B$ to $v_\perp$ {(i.e., $\dot B$ `leads')}.

To obtain a more general view of the interaction between magnetic fluctuations (measured by the Mirnov pick-up coil) and the perpendicular flow velocity  (associated with Zonal Flows), we analyze Doppler Reflectometry data from a series of discharges in which the radial position was varied systematically~\cite{Happel:2011,Estrada:2012b}; the same data have been analyzed before to calculate the auto-bicoherence~\cite{Milligen:2013}.
We do this analysis for the two of the most populated configurations in the database, namely 100\_42\_64 ($\iotabar(\rho=2/3) = 1.568$, {`fast'}) and 100\_35\_61 ($\iotabar(\rho=2/3) = 1.493$, {`slow'}).
In these two configurations, the Doppler Reflectometer measurement positions (in the H or I phase) are roughly indicated in Fig.~\ref{Config_all} by two rectangles.
Due to the expansion of the density profile (cf.~Fig.~\ref{TS}), the radial position of the Doppler Reflectometer channels prior to the transition (in the L phase) is somewhat further inward~\cite{Happel:2011}.
In configuration 100\_35\_61 ($\iotabar(\rho=2/3) = 1.493$, {`slow'}), the transition is called an L--I transition, where the I phase is characterized by limit cycle oscillations (LCOs).
In configuration 100\_42\_64 ($\iotabar(\rho=2/3) = 1.568$, {`fast'}), the transition is an L--H transition, without intermediate I phase.

\clearpage
\subsection{Spatio-temporal {analysis}, configuration 100\_35\_61 ({`slow'} transitions)}

\begin{figure}\centering
  \includegraphics[trim=0 0 400 0,clip=,width=14cm]{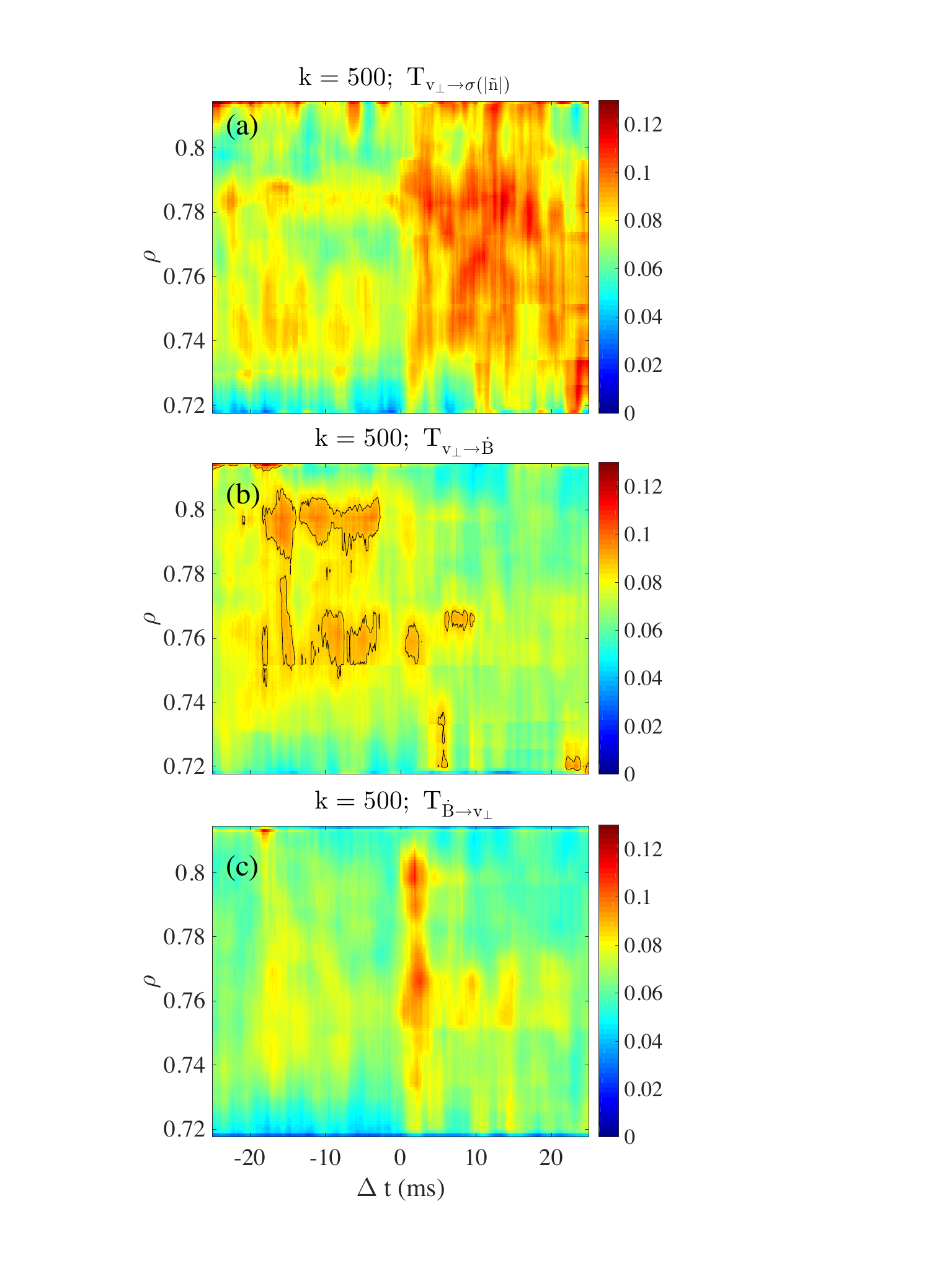}
\caption{\label{100_35}Configuration 100\_35\_61 ($\iotabar(\rho=2/3) = 1.493$).
(a) Transfer Entropy $T_{v_\perp \to \sigma(|\tilde n|)}$.
(b) Transfer Entropy $T_{v_\perp \to \dot B}$.
(c) Transfer Entropy $T_{\dot B \to v_\perp}$.
}
\end{figure}

Fig.~\ref{100_35}(a) shows the Transfer Entropy $T_{v_\perp \to \sigma(|\tilde n|)}$ ($m=5$, lag $k=500$, or 50 $\mu$s),
reflecting the interaction between the perpendicular flow velocity and the density fluctuation amplitude, both measured by the reflectometer.
{The statistical error level of the reported $T$ values is 0.01.}
$\sigma(|\tilde n|)$ is the running RMS of the density fluctuation amplitude signal {(calculated using a running time window of 2 $\mu$s)}, considered a measure of the turbulence amplitude envelope. 
The radii calculated for the Doppler Reflectometer channels vary according to the evolving density profile. 
The radii reported on the ordinate of the graphs were calculated for density profiles corresponding to $\Delta t > 0$.
The figure shows that the perpendicular velocity mainly has a causal impact {(in the restricted sense explained above)} on the density fluctuations in a period of about 30 ms after the L--I transition, in a specific radial range.
In vacuum, the radial position of the $\iotabar = 3/2$ rational is located at $\rho \simeq 0.73$ (cf.~Fig.~\ref{iota}).
Probably, the rational surface is shifted outward somewhat in the presence of the plasma with a small negative net current, and possibly coincides with the radial position at which the Transfer Entropy is showing a strong response ($\rho \simeq 0.74-0.79$).
We note that the locations and times of high Transfer Entropy coincide with the observation of LCOs in these same discharges~\cite{Estrada:2011}, i.e.,
mainly after the L--I transition. 
The reverse interaction, $T_{\sigma(|\tilde n|) \to v_\perp}$ (not shown), is much smaller, in line with the expectations based on a similar analysis for a simplified predator-prey model~\cite{Milligen:2014}, in which the dominant {direction of the} interaction between turbulence amplitude and sheared flow was found to be from sheared flow to turbulence amplitude.
Thus, these results constitute a validation of the analysis method in the framework of confinement transitions and LCOs.

As this interaction occurs mainly after the transition, {it is unlikely to be responsible for the transition.}
In the following, we {will examine the role played by magnetic fluctuations in the transition.}

Fig.~\ref{100_35}(b) shows the Transfer Entropy $T_{v_\perp \to \dot B}$ ($m=5$, lag $k=500$, or 50 $\mu$s),
to study the interaction between the perpendicular flow velocity and the magnetic fluctuations, measured by a pickup coil.
The Transfer Entropy is now only large in a time period of about 25 ms preceding the transition time.
Thus, one may presume that the interaction between $v_\perp$ and $\dot B$ is {associated with the gradual reduction of} RMS$(\dot B)$ {near} the transition, as shown in Fig.~\ref{mirnov}.

We draw attention to the fact that there seem to be two predominant zones of interaction: one at $\rho \simeq 0.76$ and one at $\rho \simeq 0.8$.
To emphasize this fact, a single level contour has been drawn in the figure (black). 
It should be noted that the radii in the L phase ($\Delta t < 0$) are somewhat smaller than indicated on the ordinate axes of the graphs (valid for the I phase, $\Delta t > 0$), as mentioned above: $\rho_L \simeq -0.656 + 1.77\rho_I$, so the actual radii are, respectively, $\rho_L \simeq 0.69$ and $0.76$, bracketing the theoretical vacuum position of the rational surface ($0.73$).
This is consistent with results from resistive MHD turbulence simulations, performed for similar parameters~\cite{Milligen:2016}.
In these simulations, the mean poloidal flow was typically found to be symmetric around a rational surface, showing a double peak inside and outside from the rational surface, at distances compatible with the two radial zones of interaction found here.

Further evidence of the interaction between velocity oscillations and magnetic fluctuations is shown in Fig.~\ref{100_35}(c), showing 
$T_{\dot B \to v_\perp}$ ($m=5$, lag $k=500$, or 50 $\mu$s).
The Transfer Entropy exhibits a short burst at the L--I transition time, at the same two radial locations as in  Fig.~\ref{100_35}(b).
This phenomenon corresponds to the short-lived interaction described in the preceding section.
We note that the direction of the {interaction} is reversed; here, the magnetic fluctuations are {`influencing'} the perpendicular {fluctuating} velocity, precisely at the transition.
{The results shown seem to indicate that magnetic fluctuations play an important role in confinement transitions.}

We note that the definition of the `transition time' used to define $\Delta t = 0$ is based on global discharge parameters such as the decay of $H_\alpha$ emission, which means that the definition may deviate from the `true' transition time by {about a ms}.
{Thus, it may be that the burst of Transfer Entropy, $T_{\dot B \to v_\perp}$, is a more precise marker of the `actual' transition.}

\clearpage
\subsection{Spatio-temporal {analysis}, configuration 100\_42\_64 ({`fast'} transitions)}

In this configuration, the plasma makes a rapid transition from the L to the H state.

\begin{figure}\centering
  \includegraphics[trim=0 0 300 0,clip=,width=14cm]{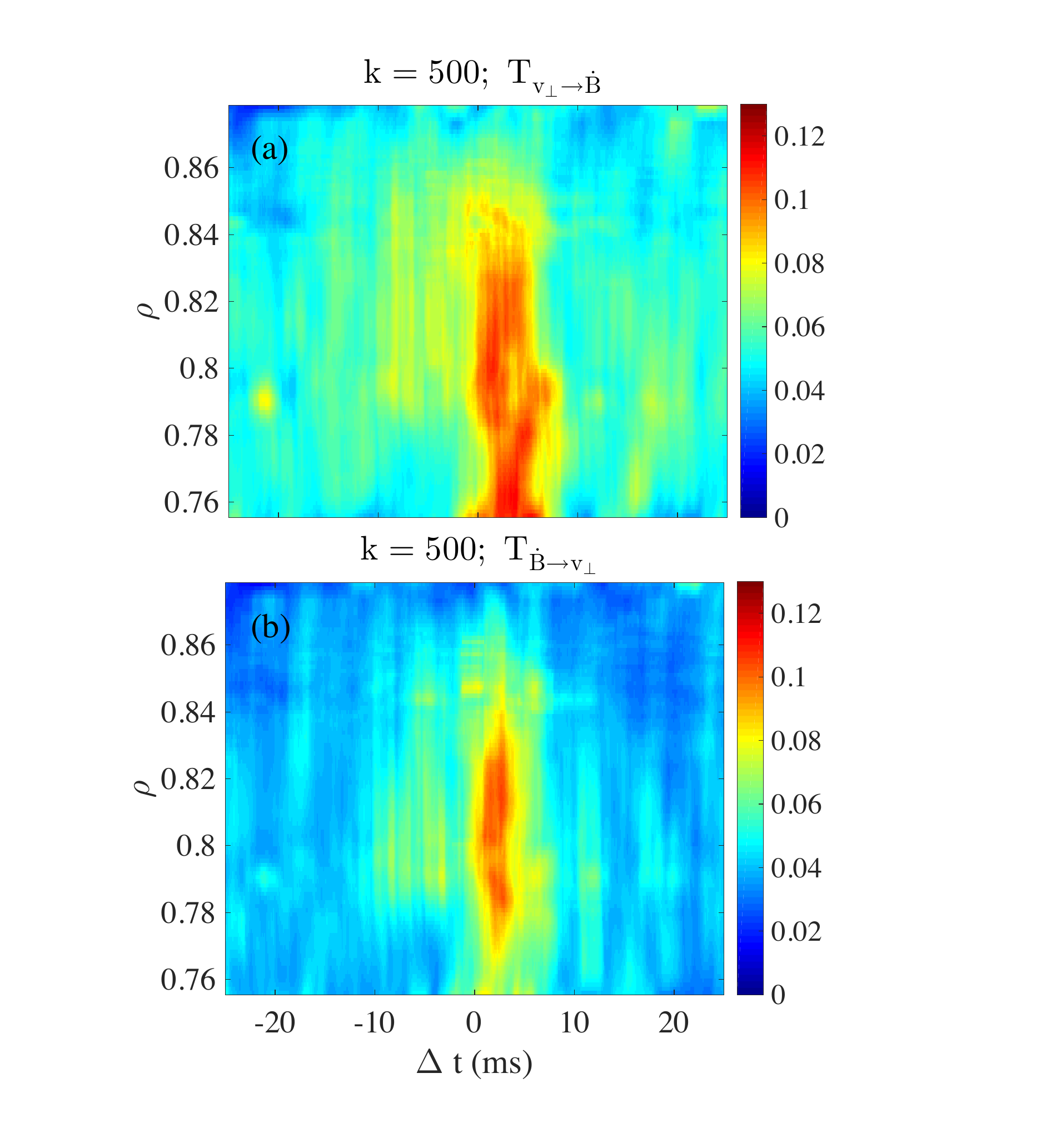}
\caption{\label{101_42}Configuration 100\_42\_64 ($\iotabar(\rho=2/3) = 1.568$).
(a) Transfer Entropy $T_{v_\perp \to \dot B}$.
(b) Transfer Entropy $T_{\dot B \to v_\perp}$.
}
\end{figure}

Fig.~\ref{101_42}(a) shows the Transfer Entropy $T_{v_\perp \to \dot B}$ ($m=5$, lag $k=500$, or 50 $\mu$s), while 
Fig.~\ref{101_42}(b) shows the Transfer Entropy $T_{\dot B \to v_\perp}$ ($m=5$, lag $k=500$, or 50 $\mu$s).
{The statistical error level of the reported $T$ values is 0.01.}
The radii shown correspond to the H phase.
The Transfer Entropy is sharply concentrated at the L--H transition.
The temporal sharpness of the interaction between perpendicular velocity and magnetic fluctuations is likely related to the sharp decay of RMS($\dot B$) shown in Fig.~\ref{mirnov} for this configuration.
In vacuum, the radial position of the $\iotabar = 8/5$ rational is located at $\rho \simeq 0.86$ (cf.~Fig.~\ref{iota}).
Probably, the rational surface is shifted inward somewhat in the presence of the plasma with a small positive net current, and possibly coincides with the radial position at which the Transfer Entropy is showing a response ($\rho \simeq 0.76-0.84$). 

Due to the fact that this transition is `fast' (as mentioned above), the sequence of interactions at the transition is not fully resolved; in particular, the sharp peak near $\Delta t \simeq 0$ seen in Fig.~\ref{23029} is not clearly visible in these {graphs that result from averaging over several discharges}.

\clearpage
\section{Discussion}\label{discussion}

In this work, we have shown how various global parameters evolve systematically across the L--H transition at TJ-II in a number of magnetic configurations.

First, we have shown that the confinement transitions are affected by the presence of rational surfaces.
The line average density at which the L--H (or L--I) transition is triggered tends to be lower when a low order rational surface is present in the edge (gradient) region (approximately, $0.6 < \rho < 0.9$), although this effect is only clear for the lowest order rationals studied ($\iotabar = n/m=3/2$ and $8/5$); cf.~Fig.~\ref{Config_all}~\cite{Wagner:2005,Estrada:2010}.

Second, we have made the observation that low frequency MHD activity is strongly suppressed {near} the L--H (or L--I) transition, cf.~Fig.~\ref{mirnov}.
The suppression is either fast (within a few ms) or slow (starting around 10 ms before the transition {time, as defined above}), depending on the magnetic configuration.

We have used a causality detection technique, {based on } the Transfer Entropy ($T$), to study the interaction between magnetic fluctuations, $\dot B$ (measured by a poloidal field pick-up coil), and the local perpendicular plasma rotation velocity, $v_\perp$ (measured by Doppler Reflectometry). 
We obtain a clear and significant interaction between magnetic fluctuations and {the fluctuating} velocity.
Making a distinction between two magnetic configurations (corresponding to different low-order rational surfaces in the edge region), we find the following.

Configuration 100\_35\_61 ($\iotabar(\rho=2/3) = 1.493$, rational $3/2$ in the edge region) is characterized by `slow' transitions and LCOs following the L--I transition.
The Transfer Entropy $T_{v_\perp \to \sigma(|\tilde n|)}$ measures the causal impact of the poloidal velocity (Zonal Flow) on the turbulence amplitude envelope, and is large after the transition, during the LCOs reported in earlier work~\cite{Estrada:2011}, cf.~Fig.~\ref{100_35}(a). 
The direction of the dominant {interaction} was in line with expectations {obtained} from a simplified predator-prey model~\cite{Milligen:2014}.

To clarify the interaction between Zonal Flows and magnetic oscillations, we calculated the Transfer Entropy $T_{v_\perp \to \dot B}$, and found that it is large at two radial positions associated with the rational surface and during several tens of ms before the transition, cf.~Fig.~\ref{100_35}(b).
Thus, it seems that the gradual reduction of magnetic fluctuation amplitude prior to the transition, reported in Fig.~\ref{mirnov}, can be explained by the gradual development of Zonal Flows associated with the rational surface.
At the transition itself, however, there is a short burst of {interactions} in the opposite direction such that $T_{\dot B \to v_\perp}$ is large, cf.~Fig.~\ref{100_35}(c), suggesting that the magnetic fluctuations play an important role during the L--I transition, { when the mean sheared flow is formed}.

In configuration 100\_42\_64 ($\iotabar(\rho=2/3) = 1.568$, rational 8/5 in the edge region), characterized by `fast' transitions without LCOs, 
$T_{v_\perp \to \dot B}$ and $T_{\dot B \to v_\perp}$ are both large in a relatively narrow time window around the transition time, and near the radial position of the rational surface, cf.~Fig.~\ref{101_42}, although the precise sequence of events is hard to follow for these `fast' transitions.
In any case, it seems the magnetic fluctuations play an essential role in the transition.

It is still unclear why the rational $8/5$ leads to a `fast' transition and the rational $3/2$ to a `slow' one, although the position and width of the islands associated with the corresponding rational and the local density gradient may play a role (cf.~Fig.~\ref{ne_profiles}). This question is left to future {work.}

\clearpage
\section{Conclusions}\label{conclusions}

In this work we have analyzed an extensive database of confinement transitions at the TJ-II stellarator.
The mean line average density at the L-H (or L--I) transition was found to be lower when a low-order rational surface was present in the edge region, cf.~Fig.~\ref{Config_all}.
Also, it was found that low frequency MHD activity was systematically and strongly suppressed {near} the transition, cf.~Fig.~\ref{mirnov}.
Together, these are clear indications that the presence of low-order rational surfaces in the region where the Zonal Flow forms (the edge or density gradient region) affects the confinement transition.

We applied a causality detection technique to reveal the detailed interaction between magnetic fluctuations (measured by a pick-up coil) and the perpendicular flow velocity (measured by Doppler Reflectometry).
We conclude that with `slow' transitions, the developing Zonal Flow prior to the transition {is associated with} the gradual reduction of magnetic oscillations.
Apparently, this reduction of the amplitude of the MHD mode oscillations is a prerequisite for the confinement transition to occur.
At the transition, we observe a strong spike of `information transfer' from magnetic to velocity oscillations, suggesting that the magnetic drive 
{may play a role in setting} up the final sheared flow, responsible for the transport barrier.
Similar observations were made for the `fast' transitions, although the temporal resolution was insufficient to clarify the sequence of events fully.

This work clearly suggests that magnetic oscillations associated with rational surfaces play an important and active role in confinement transitions,
so that electromagnetic effects should be included in any complete transition model~\cite{Scott:2005}.
Further work to confirm or study this issue on other devices is suggested.
In this framework, we consider that {causality detection based on} the Transfer Entropy constitutes an indispensable tool.

\section*{Acknowledgements}
Research sponsored in part by the Ministerio de Econom\'ia y Competitividad of Spain under project Nrs.~ENE2012-30832, ENE2013-48109-P and ENE2015-68206-P.
This work has been carried out within the framework of the EUROfusion Consortium and has received funding from the Euratom research and training programme 2014-2018 under grant agreement No 633053. 
The views and opinions expressed herein do not necessarily reflect those of the European Commission.

\clearpage
\section*{References}


\end{document}